\documentclass[final,3p,times,twocolumn]{elsarticle}
\usepackage{graphicx}
\graphicspath{ {./images/} }
\usepackage{amssymb}
 \usepackage{amsthm}
\usepackage{ulem}
\usepackage{xcolor}
\usepackage{mathtools}
\usepackage{dblfloatfix}
\usepackage{float}
\usepackage{dblfloatfix}
\usepackage{hyperref}
\journal{Journal of Magnetism and Magnetic Materials}

\begin{document}

\begin{frontmatter}


\title{Spin-singlet dimer phase in a frustrated square lattice under a magnetic field}



\author[label1]{L. M. Ramos}
\ead{lucas.morais@ufms.br}
\author[label2]{M. Schmidt}
\ead{mateus.schmidt@ufsm.br}
\author[label1]{F. M. Zimmer}
\ead{fabio.zimmer@ufms.br}
\address[label1]{Instituto de F\'{\i}sica - Universidade Federal de Mato Grosso do Sul, Campo Grande, 79070-900, MS, Brazil}
\address[label2]{Departamento de F\'{\i}sica - Universidade Federal de Santa Maria, Santa Maria,  97105-900, RS, Brazil}

\begin{abstract}
We investigated the isotropic spin-\(\frac{1}{2}\) Heisenberg model on an anisotropic square lattice with competing exchange interactions, motivated by the unconventional magnetic behavior observed in the verdazyl-based compound ($o$-MePy-V)PF$_6$. 
Using a cluster mean-field approach, we explore a field-induced phase stabilized by the interplay between frustration and quantum fluctuations, focusing on the role of exchange interactions.
We identify: (i) the formation of spin singlet pairs, signaled by enhanced spin–spin correlations in specific field regimes; and (ii) a one-half magnetization plateau, emerging from a subtle balance between competing exchange couplings and field-enhanced quantum fluctuations. 
Our results reveal that an enhancement of frustration, achieved by tuning small variations in the spatially anisotropic exchange interactions of the compound ($o$-MePy-V)PF$_6$, can stabilize a field-induced quantum phase where ferromagnetism coexists with antiferromagnetic dimers.
Our results provide microscopic insight into the mechanisms driving these nontrivial phases and offer theoretical support for interpreting experimental observations in this class of low-dimensional quantum magnets.
\end{abstract}


\end{frontmatter}

%\linenumbers

\section{\label{sec:level1}Introduction}

Magnetic systems have long served as fertile ground for uncovering emergent quantum phenomena, especially when competing interactions create a rich landscape of degenerate ground states and excitations \cite{Savary_2017,Zhou}.
For instance, the combination of competing interactions, low dimensionality and strong quantum fluctuations can stabilize a variety of unconventional phases such as spin liquids \cite{Jiang,Metavitsiadis,LIU20221034}, valence bond solids \cite{Xi}, and fractional magnetization plateaus \cite{LiHZ}. 
These states of matter are not only of fundamental interest but also provide insights into entanglement \cite{Laflorencie,Song}, quantum criticality \cite{Yang}, and exotic excitations such as spinons \cite{Tran,Shao} and triplons \cite{Honecker,Udono}.
Moreover, tuning quantum fluctuations via an external magnetic field in frustrated systems can lead to intriguing effects \cite{Yamaguchi, Yamaguchi2021}, such as the coexistence of magnetization plateaus and dimerized states \cite{lucas}.
An important question concerns whether the degree of frustration can play a central role in the stability of such states. In the present work, we investigate this issue on a spin-1/2 frustrated square lattice with anisotropic interactions that hosts a spin-singlet state.

Frustration can arise in a variety of systems, either from the lattice geometry or due to the nature and range of the interactions. Typical candidates for exhibiting frustration include triangular-based lattices with antiferromagnetic (AF) exchange interactions \cite{ 
10.1063/1.2186278,wang2025}, disordered systems such as spin glasses \cite{Mydosh_2015,Eric, Schmidt_2017, Zimmer2014}, and even bipartite lattices when there is competition between nearest- and next-nearest-neighbor interactions or between AF and ferromagnetic (FM)  couplings \cite{10.1093/ptep/ptae061, LIU20221034}.
In particular, the presence of  FM and AF couplings on different bonds can introduce a subtle competition, in which a delicate balance of interactions can favor ground states with nontrivial spin arrangements in bipartite systems \cite{PhysRevB.103.L220407}. 
A prototypical example is found in spin-ladder systems, where AF interchain (rung) interactions compete with  FM intrachain (leg) couplings, giving rise to a rich variety of unconventional magnetic phases, including partially ordered states, spin-singlet dimer phases, and magnetization plateaus under applied magnetic fields \cite{Liu,wessel2017efficient,almeidabibiano,Saito}.

Another interesting example of competitive system is found in the verdazyl-based molecular compound ($o$-MePy-V)PF$_6$, which provides a compelling platform for studying frustrated quantum magnetism \cite{Yamaguchi}. 
This compound features a quasi-two-dimensional spin-1/2 square lattice, with PF$_6$ anions intercalated between the layers, and exhibits a collinear AF ground state with twofold periodicity at zero magnetic field. 
However, its magnetization under an applied magnetic field deviates from conventional behavior, exhibiting an increase with alternating concavity, featuring a plateau-like structure at low temperatures.
This outcome is well described by an anisotropic-lattice Heisenberg model \cite{Yamaguchi, lucas}.
Remarkably, when the model also includes a weak spin anisotropy, the plateau-like behavior can evolve into a well-defined magnetization plateau. 
More importantly, the interplay between frustration and field-induced quantum fluctuations brings out the coexistence of  FM and dimerized chains in this plateau regime \cite{lucas}.

It has been found that spin anisotropy can help stabilize this exotic state, which is absent in the isotropic spin regime. However, it is still unclear whether this dimerized state can also be accessed in the isotropic limit of the model through small changes in the exchange interactions proposed in Ref. \cite{Yamaguchi}.
Therefore, a natural question is whether tuning the exchange interactions and, thereby, the degree of frustration in the spin-1/2 Heisenberg model can lead to a field-induced dimerized state on the square lattice inspired by the compound ($o$-MePy-V)PF$_{6}$.

In this work, we present a theoretical investigation of the spin-1/2 Heisenberg model on a square lattice with six distinct nearest-neighbor exchange interactions under an external magnetic field.
Our study focuses on how the competing exchange interactions and applied magnetic field can contribute to stabilizing novel quantum phases, with particular emphasis on:
(i) the singlet-pair correlations, quantified through spin-spin correlation functions, and
(ii) the emergence of one-half magnetization plateau, arising from the subtle balance between the degree of frustration and enhanced quantum fluctuations. 
We investigate the ground-state properties of this frustrated model by employing the cluster mean-field (CMF) method. 
This approach improves the conventional mean-field theory by incorporating exact calculations of the short-range correlations within finite-size clusters, while treating intercluster couplings at the mean-field level. 
This method has demonstrated remarkable accuracy in capturing essential features of frustrated quantum magnets at low temperatures, including magnetization plateaus, nontrivial quantum phase transitions, and nonmagnetic states that arise from the competition between exchange interactions and quantum fluctuations \cite{yamamoto2014, yamamoto2015, yamamoto2017, Singhania, kellermann2019quantum}. 
The CMF method enables the identification of quantum phases and field-induced phenomena \cite{lucas}, offering quantitative agreement with previous numerical studies based on the tensor network method \cite{Yamaguchi} and focused on the model proposed for the compound ($o$-MePy-V)PF$_{6}$.

After introducing the model in more detail in the following Sec. \hyperref[sec:model]{2}, we present the CMF results for the ground state and the effects of thermal fluctuations in Sec. \hyperref[sec:results]{3}. 
Finally, Sec. \hyperref[sec:summary]{4} summarizes the paper and presents our conclusion.

\section{Model and method}\label{sec:model}

\begin{figure}[!t]
    \centering    \includegraphics[width=.4\textwidth]{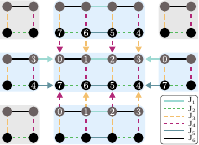}
    \caption{Schematic representation of the eight-site CMF approach applied to an anisotropic square lattice with six distinct exchange couplings \(J_{n}\) (\(n = 1,\dots,6\)). Solid and dashed lines denote  FM and AF exchange interactions, respectively. The gray sites form a  FM chain connected by couplings \(J_{1}\) and \(J_{6}\) (referred to as chain 1), while the black sites form chain 2, composed of AF and  FM couplings \(J_{2}\) and \(J_{5}\). The two chains are coupled via AF interactions \(J_{3}\) and \(J_{4}\), resulting in a frustrated square lattice geometry. Frustration arises in square plaquettes containing an odd number of AF bonds. Arrows indicate the bonds treated at the mean-field level within the CMF framework.}
    \label{fig:CMF}
\end{figure}

We consider the spin-\(\frac{1}{2}\) Heisenberg model on an anisotropic square lattice with six distinct nearest-neighbor exchange couplings \(J_{n}\) (\(n = 1, \dots, 6\)). The Hamiltonian is given by
\begin{equation}\label{eq:ham}
H = -\sum_{\langle i,j \rangle} J_n \, \mathbf{S}_i \cdot \mathbf{S}_j - h^z \sum_i S^z_i,
\end{equation}
where \(\mathbf{S}_{i}\) denotes the spin-\(\frac{1}{2}\) operators at site \(i\), and \(J_{n}\) represents the anisotropic exchange couplings between neighboring sites \(i\) and \(j\). The last term in Eq.~\ref{eq:ham} represents the applied magnetic field $h^z$, which acts along the \(S^{z}\) spin component.  
The parametrization of \(J_n\) follows a similar pattern of signs to that adopted in the model proposed in Ref.~\cite{Yamaguchi} and is schematically illustrated in Fig.~\ref{fig:CMF}.

The lattice structure can be viewed as composed of two coupled spin chains (see Fig.~\ref{fig:CMF}). Chain 1 (gray sites) is a  FM chain governed by couplings \(J_1\) and \(J_6\), while chain 2 (black sites) consists of a mixed AF–FM chain with AF \(J_2 < 0\) and  FM \(J_5 > 0\) couplings. The two chains are connected via AF interactions \(J_3\) and \(J_4\), forming an anisotropic and frustrated square lattice. Frustration arises in square plaquettes containing an odd number of AF bonds, characterized by loops for which the product of exchange couplings satisfies \(\prod_n J_n < 0\). In particular, loops involving \(J_2\) are frustrated.

To investigate this system, we employ the CMF method, in which the infinite lattice is partitioned into identical clusters of eight sites, \(n_s = 8\), as illustrated by the shaded regions of Fig. \ref{fig:CMF}. 
The intercluster interactions are decoupled according to
\begin{equation}\label{eq:mean}
\mathbf{S}_i \cdot \mathbf{S}_j \approx \mathbf{S}_i \cdot \langle \mathbf{S}_j \rangle + \langle \mathbf{S}_i \rangle \cdot \mathbf{S}_j - \langle \mathbf{S}_i \rangle \cdot \langle \mathbf{S}_j \rangle,
\end{equation}
where $\langle \mathbf{S}_i \rangle$ represents the local magnetization at site \(i\).
This decoupling leads to the effective single-cluster Hamiltonian
\begin{equation}\begin{split}\label{eq:hcmf}
H_{\mathrm{eff}} = & -\sum_{\langle i,j \rangle} J_n \, \mathbf{S}_i \cdot \mathbf{S}_j - h^z \sum_i S^z_i \\
& -\sum_{(i,k)} J_{\bar{n}} \sum_{\alpha} \left[ S_i^\alpha \langle S_k^\alpha \rangle - \frac{1}{2} \langle S_i^\alpha \rangle \langle S_k^\alpha \rangle \right],
\end{split}\end{equation}
where \((i,k)\) labels pairs of boundary sites in neighboring clusters coupled via \(J_{\bar{n}}\), with \( \langle S^{\alpha} \rangle\) denoting the local magnetization along the direction \(\alpha = x, y, z\).
In the present approach, we consider that the system exhibits identical patterns of local magnetizations across all clusters, which allows the local magnetizations of neighboring clusters \(\langle S^{\alpha}_{k} \rangle\) to be obtained from the corresponding sites within the considered cluster.
In this case, the summation $\sum_{(i,k)}$ in Eq. (\ref{eq:hcmf}) can be evaluated by replacing $\langle S_{k}^{\alpha}\rangle$ with the corresponding local magnetization pattern at the site of the considered cluster. For example, when $i=0$, the sum term \(J_{\bar{n}}S_i^{\alpha}\langle S_{k}^{\alpha}\rangle\) is replaced by \(J_4S_0^{\alpha}\langle S_{7}^{\alpha}\rangle+J_6S_0^{\alpha}\langle S_{3}^{\alpha}\rangle\), where all the sites belong to the considered cluster.

The local magnetizations are determined self-consistently from the thermal expectation values
\begin{equation}\label{eq:local_mag}
m_i^\alpha \equiv\langle S^{\alpha}_{i}\rangle= \frac{\mathrm{Tr} \left( S_i^\alpha e^{-\beta H_{\mathrm{eff}}} \right)}{Z}, \quad 
Z = \mathrm{Tr} \left( e^{-\beta H_{\mathrm{eff}}} \right),
\end{equation}
where \(\beta = 1/(k_B T)\) and \(Z\) is the partition function of the cluster. In the zero-temperature limit, the expectation values reduce to the ground-state expectation values over the lowest-energy eigenstate.

The self-consistent solution corresponds to the stationary point that minimizes the variational free energy associated with the CMF ansatz. Specifically, the free energy per cluster is given by $F = -\frac{1}{\beta} \ln Z$.  The physically stable solution corresponds to the minima of \(F\) with respect to the set of local magnetizations \(\{ m_i^\alpha \}\). This guarantees thermodynamic consistency within the CMF approximation.

Once the self-consistent solution is obtained,  other thermodynamic quantities can be computed. The total magnetization is evaluated as
\begin{equation}\label{eq:mtotal}
M = \sqrt{ (M^x)^2 + (M^y)^2 + (M^z)^2 }, \quad 
M^\alpha = \frac{1}{n_s} \sum_{j} m_j^\alpha,
\end{equation}
and in order to characterize the magnetic properties of each chain individually, we define the average local moments
\begin{equation}\label{eq:mc1}
m_{c1}^2 = \frac{1}{4} \sum_{j=0}^{3} \sum_{\alpha} \langle S_j^\alpha \rangle^2, \quad
m_{c2}^2 = \frac{1}{4} \sum_{j=4}^{7} \sum_{\alpha} \langle S_j^\alpha \rangle^2,
\end{equation}
where sites \(j = 0\) to 3 belong to chain 1 and sites \(j = 4\) to 7 belong to chain 2.
These quantities measure the average magnitude of the local spins within each chain, rather than their alignment (i.e., the sign of the local magnetizations is irrelevant for these quantities).
Furthermore, the spin–spin correlation functions are calculated to probe the nature of quantum correlations and the presence of short-range order:
\begin{equation}\label{corrSiSj}
C_{ij} = \langle \mathbf{S}_i \cdot \mathbf{S}_j \rangle.
\end{equation}

This procedure indicates that the CMF method provides a versatile framework for investigating various physical quantities, with intracluster dynamics fully accounted by exactly incorporating the interactions within each cluster. 
The intercluster interactions are treated self-consistently to capture local features of the model. 
In this sense, the CMF approach offers valuable insights into the interplay between quantum fluctuations, frustration, and external magnetic fields, particularly with regard to intracluster interactions. In the following section, the numerical results for the model investigated within the CMF framework are presented and analyzed.

\section{Results and discussion}\label{sec:results}

The set of local magnetizations (see Eq. (\ref{eq:local_mag})) is solved self-consistently for a cluster of up to 8 sites represented by Eq. (\ref{eq:hcmf}), which is exactly diagonalized. 
For numerical purposes, we use the exchange interaction $J_1$ as the energy unit, setting the Boltzmann constant $k_B = 1$. 
We start from the exchange couplings reported for the compound (o-MePy-V)PF$_6$, given by $J_2/J_1 = -0.82$, $J_3/J_1 = -0.66$, $J_4/J_1 = -0.61$, $J_5/J_1 = 0.26$, and $J_6/J_1 = 0.20$ \cite{Yamaguchi}, and make small variations in these in order to explore possible field-induced singlet states.
It means that the magnetic properties of the model are analyzed for different sets of exchange interactions by evaluating the Eqs. (\ref{eq:mtotal}), (\ref{eq:mc1}), and (\ref{corrSiSj}).
We first investigate the ground-state properties and then present the effects of thermal fluctuations on the magnetic correlations of the system.

\begin{figure}[!t]
    \centering
\includegraphics[width=0.48\textwidth]{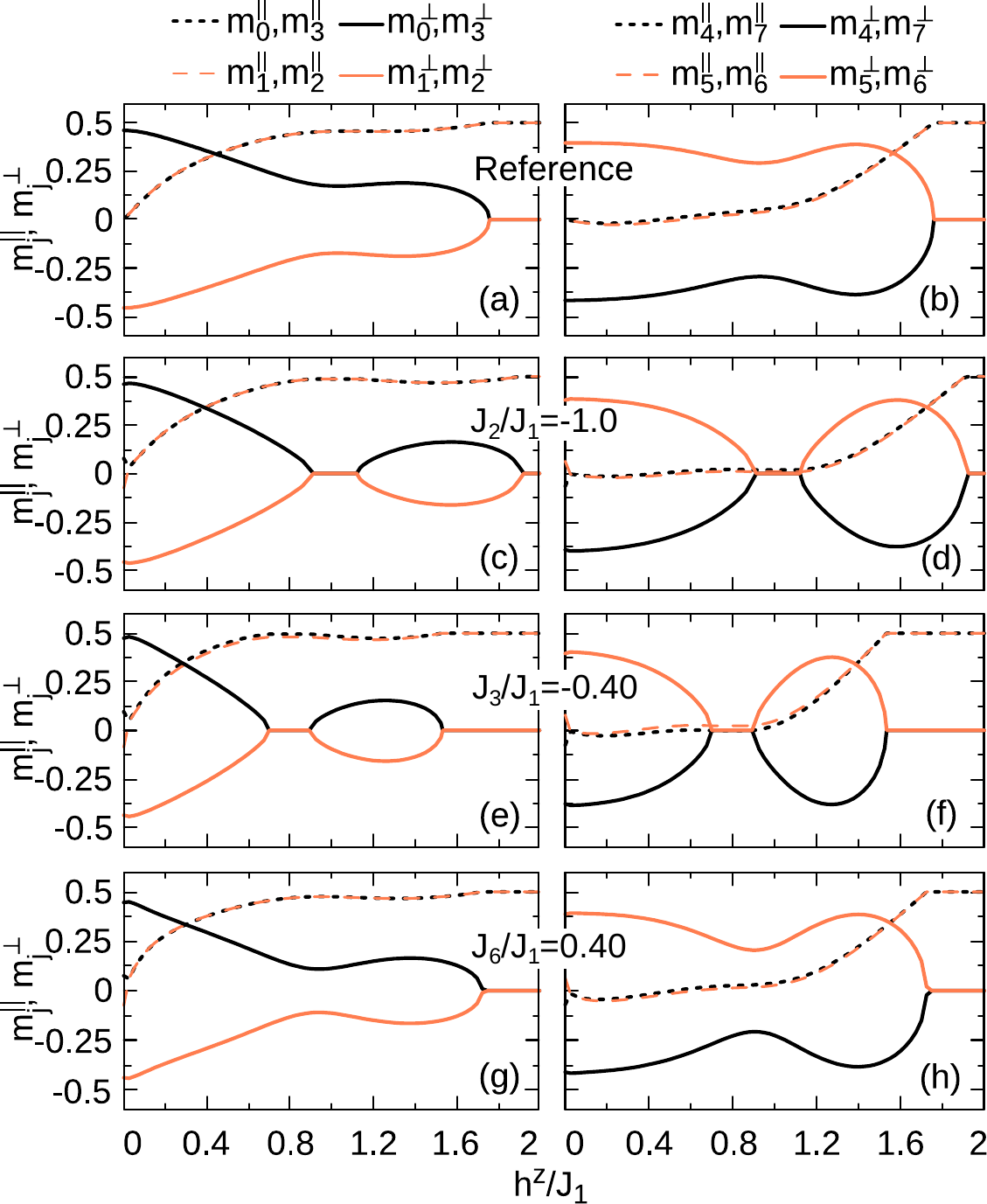}
    \caption{Longitudinal (dashed lines) and transverse (solid
     lines) components of the eight local magnetizations in the
     ground state under an external magnetic field for: (a) and (b) Reference system \label{reference}, (c) and (d) $J_{2}/ J_{1}=-1.0$\label{j2}, (e) and (f) $J_{3}/ J_{1}=-0.40$ \label{j3}, (g) and (h) $J_{6}/ J_{1}=0.40$ \label{j6}. The left column displays the local moments of chain 1, while the right column shows the local moments of chain 2.}
    \label{fig:local_mags}
\end{figure}

\subsection{Properties of the ground state}

The compound ($o$-MePy-V)PF$_6$ provides an interesting experimental perspective for studying the interplay of frustration and field-induced quantum fluctuations. It has a molecular structure composed of two quantum spin chains coupled via AF interactions. This system can be represented by an anisotropic square lattice model with six nearest-neighbor exchange interactions, in which one of the chains present both  FM and AF couplings. 
The degree of frustration in the present anisotropic square lattice model is directly associated with the strength of the exchange interactions. 
The role of frustration in the field-induced magnetic properties can be clarified by slight changes in the interactions.
Figure \ref{fig:local_mags} exhibits the local magnetization patterns for each chain with different sets of exchange interactions. This allows us to analyze the effects caused by the variation of a single exchange parameter compared to the reference compound ($o$-MePy-V)PF$_6$ shown in Fig. \ref{fig:local_mags}\hyperref[reference]{(a)} and \ref{fig:local_mags}\hyperref[reference]{(b)}.
In Figs. \ref{fig:local_mags}\hyperref[reference]{(a)} and \ref{fig:local_mags}\hyperref[reference]{(b)}, the longitudinal components vanish, whereas the transverse components are finite and antisymmetric at a zero magnetic field.
As the magnetic field increases, the transverse components of chain 1 (left panel) monotonically decrease and vanish at a critical field, while the longitudinal components smoothly grow towards saturation. 
In contrast, the longitudinal components of chain 2 (right panel) exhibit a slight and unexpected reduction at low fields, in agreement with Refs. \cite{Yamaguchi,lucas}. 
The local magnetization pattern reveals a collinear AF order within the transverse plane at zero field, characterized by a two-fold periodicity structure. 
The Zeeman coupling suppresses this AF collinear order as the field increases.

A small variation only in the exchange coupling $J_2$ (from $J_2/J_1=-0.82$ to $J_2/J_1=-1.00$) leads to a substantial modification of the local magnetization pattern at finite field, as illustrated in Figs. \ref{fig:local_mags}\hyperref[j2]{(c)} and \ref{fig:local_mags}\hyperref[j2]{(d)}. 
The longitudinal components of chain 1 present a small value at zero field, increasing smoothly to full polarization as the magnetic field increases. This dependence is comparable to that of the reference system. 
In contrast, the transverse components exhibit quite distinct behavior under an applied field. 
In both chains, the transverse components still indicate a collinear AF state in the absence of a field but gradually decrease, vanishing within a narrow field range.
In particular, chain 2 evolves to a nonmagnetic regime characterized by the absence of both transverse and longitudinal local magnetizations, while chain 1 simultaneously develops a FM ordering.
Further increasing the field leads to a continuous reemergence of these transverse components before they disappear in the saturation field.
An analogous result is obtained when exchange parameters of the original compound are used except for $J_3$, which is slightly reduced from $J_3/J_1=-0.66$ to $J_3/J_1=-0.40$ in Figs. \ref{fig:local_mags}\hyperref[j3]{(e)} and \ref{fig:local_mags}\hyperref[j3]{(f)}. 
However, changing only the exchange parameter $J_6$ (from $J_6/J_1=0.20$ to $J_6/J_1=0.40$) does not significantly affect the local magnetization pattern (see \ref{fig:local_mags}\hyperref[j6]{(g)} and \ref{fig:local_mags}\hyperref[j6]{(h)}), which essentially exhibits behavior analogous to that of the reference system.
These results suggest that small variations in the exchange parameter $J_{2}$ or $J_{3}$ can drive the system to a regime of strong quantum fluctuations induced by the field. 
In order to clarify this notable phenomenon, we will also investigate the behavior of magnetization plateaus, as well as the spin–spin correlation functions, under the effects of frustration arising from variations of such exchange parameters.

\begin{figure}[!t]
    \centering
\includegraphics[width=0.48\textwidth]{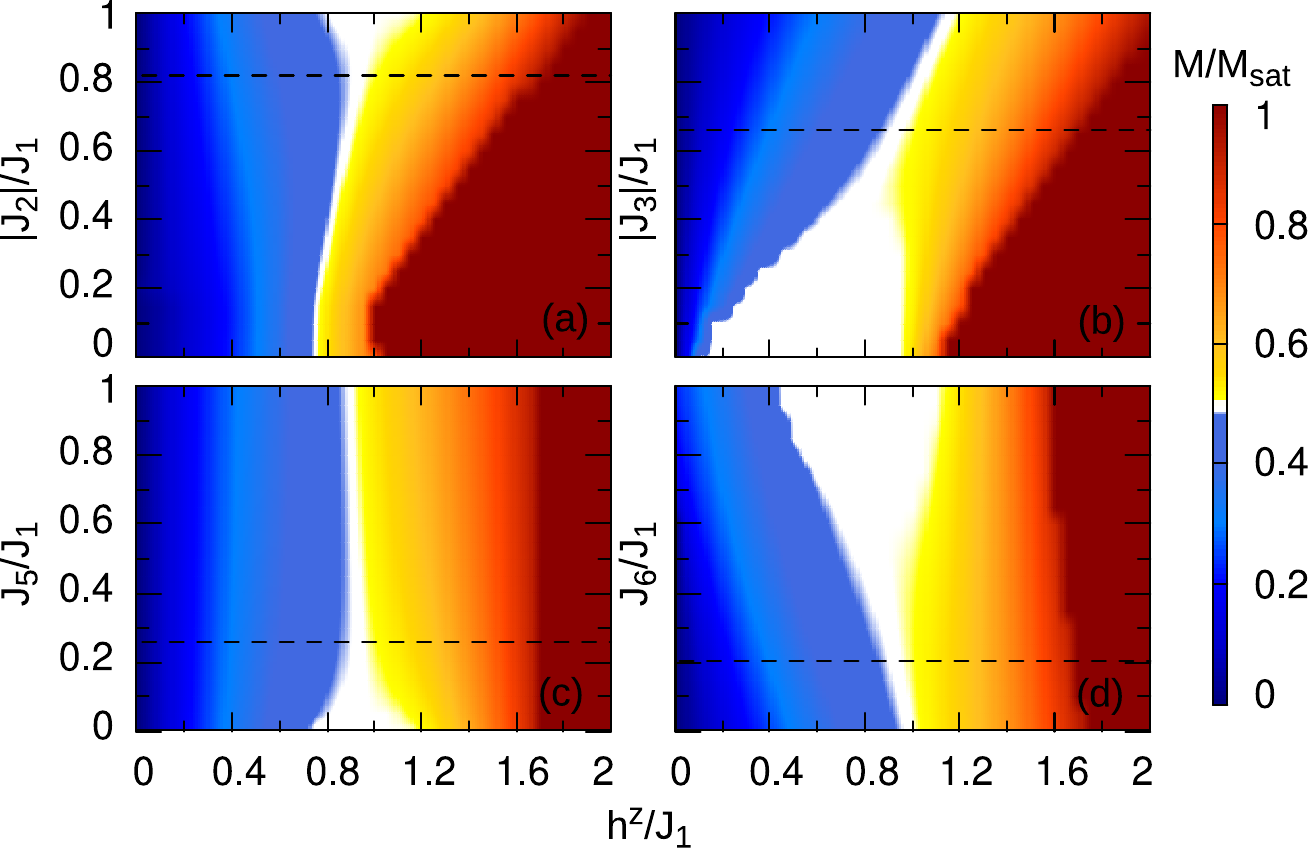}
    \caption{ Normalized total magnetization as a function of the external magnetic field, with the exchange parameters of the compound ($o$-MePy-V)PF$6$ kept fixed except for one, which is varied: (a) $|J_{2}|/J_{1}$ \label{fig:5(a)}, (b) $|J_{3}|/J_{1}$\label{fig:5(b)},  (c) $J_{5}/J_{1}$\label{fig:5(c)}, and (d) $J_{6}/J_{1}$\label{fig:5(d)}. The dashed lines represent the reference values for the exchange interactions of the compound ($o$-MePy-V)PF$_6$. 
    The one-half magnetization plateau is depicted as a broad white region in the diagram.}
    \label{fig:mags_vs_jns}
\end{figure}

The role of these exchange parameters in the field-dependence of magnetization is depicted in the heat maps of Fig. \ref{fig:mags_vs_jns}. 
Each panel of this figure shows the effect of tuning a single exchange parameter away from the reference value \cite{Yamaguchi}, which is indicated by the dashed lines.
As illustrated in Fig. \ref{fig:mags_vs_jns}\hyperref[fig:5(a)]{(a)}, a one-half magnetization plateau (depicted as a broad white region in the diagram) appears when $|J_{2}|/J_{1}$ approaches $J_1$ in the range $0.9 \lesssim h^{z}/J_{1} \lesssim 1.1$.
On the other hand, reducing  $|J_{2}|/J_{1}$ from the reference value suppresses this magnetization plateau and leads to saturation at lower magnetic field values.
Notably, the occurrence of a well-defined magnetization plateau for higher AF $J_2$ couplings suggests an intriguing regime where a nontrivial phase can be obtained.

Enhancing the AF character of chain 2 proves crucial for stabilizing the one-half plateau state. 
This is supported by Fig. \ref{fig:mags_vs_jns}\hyperref[fig:5(c)]{(c)}, where weakening the  FM $J_5$ coupling in chain 2 expands the magnetization plateau region.
However, our previous analysis is only valid when the chains are coupled by AF exchange interactions,  which can lead to frustration.
For example, Fig. \ref{fig:mags_vs_jns}\hyperref[fig:5(b)]{(b)} shows that the AF interaction $J_3$
directly affects the total magnetization of the ground state, with a one-half plateau emerging for $|J_{3}|/J_{1}\lesssim 0.5$. We also note that variations in $J_{4}$ yield analogous results due to their equivalence with $J_{3}$ in the lattice structure.
As $|J_{3}|/J_{1}$ increases (or $|J_{4}|/J_{1}$ ), the AF character between chains becomes more pronounced, leading to a gradual suppression of the magnetization plateau.
We propose that this affects the AF spin-spin correlation related to the interaction $J_2$.

In summary, strengthening the AF character of the chain 2 enhances the degree of frustration in the square lattice, stabilizing the one-half plateau state in a regime dominated by quantum fluctuations induced by the field. 
As discussed in the following, the competition between frustration and field-induced quantum fluctuations can drive the emergence of nontrivial quantum phases within the plateau state. 
Similar mechanisms have been observed in other bipartite systems, where the interplay between frustration and quantum fluctuations leads to exotic dimerized phases \cite{pikulski2020two}. 
Finally, as expected, enhancing the  FM character of chain 1 can also be reflected in the rise in the one-half plateau state and the emergence of spontaneous magnetization at zero field (light blue color), as illustrated in Fig. \ref{fig:mags_vs_jns}\hyperref[fig:5(d)]{(d)}.
In particular, the results presented by Fig. \ref{fig:mags_vs_jns} can help classify the choice of the reference system exchange parameters as belonging to a crossover region between the well-defined magnetization plateau and a plateau-like behavior.
In summary, our results for the ground-state magnetization of this frustrated magnetic system are strongly dependent on the subtle balance between frustration and field-induced quantum fluctuations, revealing a rich phenomenology beyond the magnetism reported for the compound ($o$-MePy-V)PF$_{6}$ \cite{Yamaguchi}.

\begin{figure}[!t]
    \centering
    \includegraphics[width=0.48\textwidth]{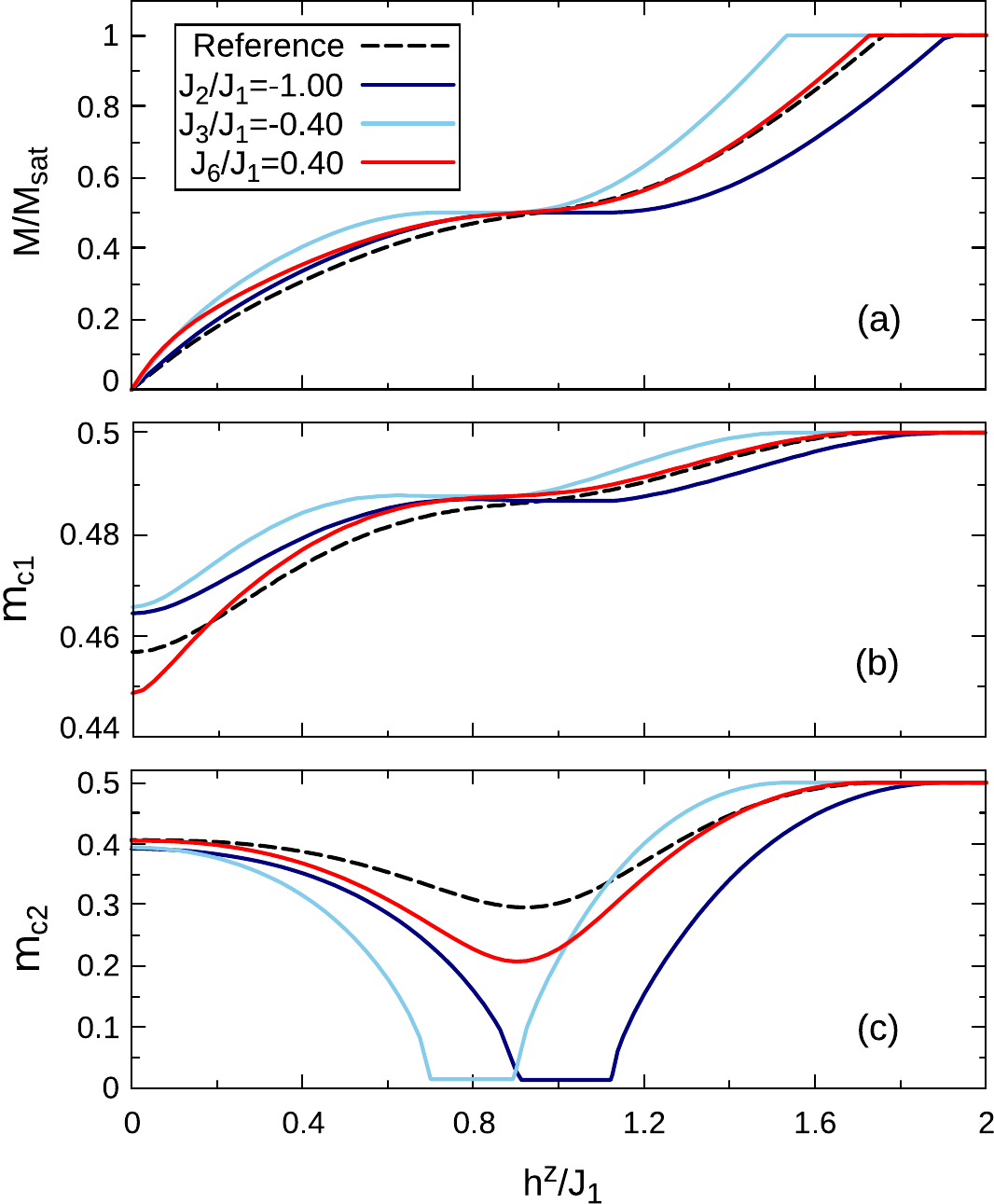}
    \caption{Ground state of the CMF model under magnetic field and different exchange interaction. (a) Total magnetization curve normalized by the saturation\label{totalmag}, (b) average local moments of chain 1 $m_{c1}$\label{mc1}, and (c) average local moments of chain 2 $m_{c2}$\label{mc2}. The dashed line indicates the exchange interactions of the reference compound ($o$-MePy-V)PF$_{6}$.}
    \label{fig:mags_hz}
\end{figure}

To account for the effects of external magnetic field on the plateau occurrence, we present in Fig. \ref{fig:mags_hz} the total magnetization and the average local moments of chains for some planes of the heat maps in Fig. \ref{fig:mags_vs_jns} (corresponding to the same exchange parameters as set in Fig. \ref{fig:local_mags}).
We begin by emphasizing that the total magnetization curve of the reference system, shown by the dashed lines in Fig. \ref{fig:mags_hz}\hyperref[totalmag]{(a)}, exhibits a plateau-like behavior that closely resembles that obtained from tensor network calculations in Ref.\cite{Yamaguchi}. Small variations in the exchange couplings relative to the reference system can either enhance this plateau-like feature or stabilize a well-defined magnetization plateau.
Our findings support that increasing the AF $J_2$ interaction (relative to the reference system parameters) enhances the degree of frustration and amplifies the field-induced quantum fluctuations. 
This effect stabilizes the magnetization plateau state, as evidenced by the dark-blue curve in Fig. \ref{fig:mags_hz}\hyperref[totalmag]{(a)}.
In this case, the average magnetic moment of chain 1 increases with field intensity but presents a constant value within the plateau range (see the dark-blue curve in Fig. \ref{fig:mags_hz}\hyperref[mc1]{(b)}). 
In contrast, chain 2 undergoes a significant reduction in the average local moment due to the applied field. For $0.9 \lesssim h^{z}/J_{1}\lesssim 1.1$, $m_{c2}$ approaches zero (see the dark-blue curve in Fig. \ref{fig:mags_hz}\hyperref[mc2]{(c)}), indicating that field-induced quantum fluctuations indeed destroy the local magnetic order.
The presence of cusps in $m_{c}$ reflects abrupt changes in the average magnitude of local moments within chain 2, such as enhanced spin fluctuations. Although the total magnetization M remains a smooth function of the magnetic field, spin-spin correlations of chain 2 can also signal this enhancement, as discussed in the following.  
These results show that the one-half plateau state for $J_{2}/J_{1}=-1.0$ is characterized by full polarization in chain 1, while chain 2 realizes a nonmagnetic phase. 
Similar coexistence of polarized and nonmagnetic regions has been reported in other frustrated spin systems, including the field-induced plateau phase of Co$_{2}$V$_{2}$O$_{7}$, where unequivalent magnetic sites exhibit distinct behaviors under an applied field \cite{LiHZ} as well as in spin-1/2 Heisenberg branched chains showing ferrimagnetic plateaus stabilized by competing interactions \cite{Karlov}. 
Related phenomena also appear in coupled trimer chain models, where fractional magnetization plateaus emerge from the interplay of polarized and singlet states \cite{montenegro}. 
We emphasize that smooth behavior of magnetization near the plateau obtained in the present study contrasts with the findings for highly frustrated systems such as the kagome lattice \cite{PhysRevB.93.060407}, which exhibits cusps in the magnetization.    

\begin{figure}[!t]
    \centering
\includegraphics[width=0.48\textwidth]{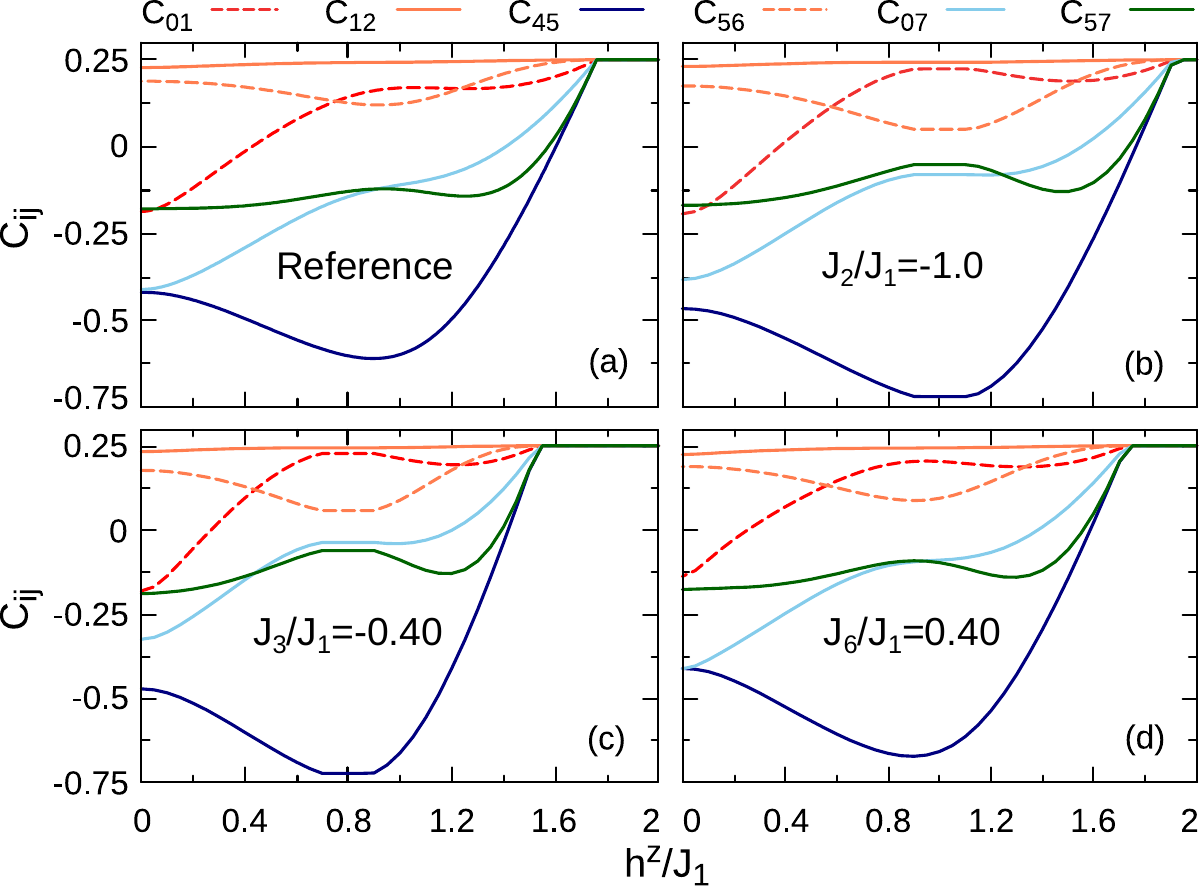}
    \caption{Field dependence of spin-spin correlation functions computed using eight-site CMF model at $T/J_{1}$=0 for (a) Reference compound  \label{fig:ref}, (b) $J_{2}/J_{1}=-1.0$\label{fig:singlet}, (c) $J_{3}/J_{1}=-0.40$\label{fig:j3} and (d) $J_{6}/J_{1}=0.40$\label{fig:j6}. Some correlations have been omitted from the figure as they are nearly equivalent to those already shown, e.g., the correlations $C_{01}\approx C_{23}$, $C_{45}\approx C_{67}$, $C_{46}\approx C_{57}$ and the interchain correlations $C_{07}\approx C_{16} \approx C_{25} \approx C_{34}$. }
    \label{fig:correlations}
\end{figure}
The coexistence of polarized states with the nonmagnetic phase can also be achieved by tuning other values of the exchange couplings. For instance, the light-blue curve in Fig. \ref{fig:mags_hz} illustrates this behavior for $J_{3}/J_{1}=-0.40$ within the field range $0.7 \lesssim h^{z}/J_{1}\lesssim 1.0$.
In addition, an analogous result to that of the reference system is recovered by setting $J_{6}/J_{1}=0.40$ (see the red curve in Fig. \ref{fig:mags_hz}), where chain 1 remains polarized along the field direction, while chain 2 exhibits a suppression of its average local moments.

The underlying physics of the previously discussed one-half magnetization plateau and nonmagnetic state in chain 2 can be further explored using spin-spin correlation functions. 
For example, Fig. \ref{fig:correlations}\hyperref[fig:ref]{(a)} illustrates the field dependence of the $C_{ij}$ for the reference system. 
At zero magnetic field, the $C_{ij}$ values for chains 1 and 2 align with a collinear AF state, where $C_{01}$, $C_{23}$, $C_{45}$, and $C_{67}$ show AF behavior, while $C_{12}$ and $C_{56}$ show  FM correlations. 
As the magnetic field increases, the correlations $C_{45}$ and $C_{67}$ show an enhanced AF character, attributed to increased quantum fluctuations.
This increase in AF character is also observed when setting $J_{2}/J_{1}=-1.0$ or $J_{3}/J_{1}=-0.40$ (see Fig. \ref{fig:correlations}\hyperref[fig:singlet]{(b)}  and \ref{fig:correlations}\hyperref[fig:j3]{(c)}), but with a fundamental difference: the correlations $C_{45}$ and $C_{67}$ approach the perfect singlet value of \textminus0.75, while the interchain correlations (e.g., $C_{07}$, $C_{16}$,...) and $C_{56}$ approach zero.
Furthermore, the correlations $C_{46}$ and $C_{57}$ are also small, suggesting that chain 2 effectively forms singlet pairs and becomes dimerized due to the absence of local magnetic order (see Fig. \ref{fig:local_mags}\hyperref[j2]{(d)} and \ref{fig:local_mags}\hyperref[j3]{(f)}).
This indicates that the one-half plateau state is effectively characterized by  FM ordering in chain 1 and the realization of a spin-singlet dimer phase in chain 2.
By changing only $J_{6}$ (from $J_6/J_1=0.20$ to $J_6/J_1=0.40$) in Fig. \ref{fig:correlations}\hyperref[fig:j6]{(d)}, the formation of singlet pairs described previously is absent, and the correlation pattern is similar to that of the reference compound.
\begin{figure}[!t]
    \centering
    \includegraphics[width=0.48\textwidth]{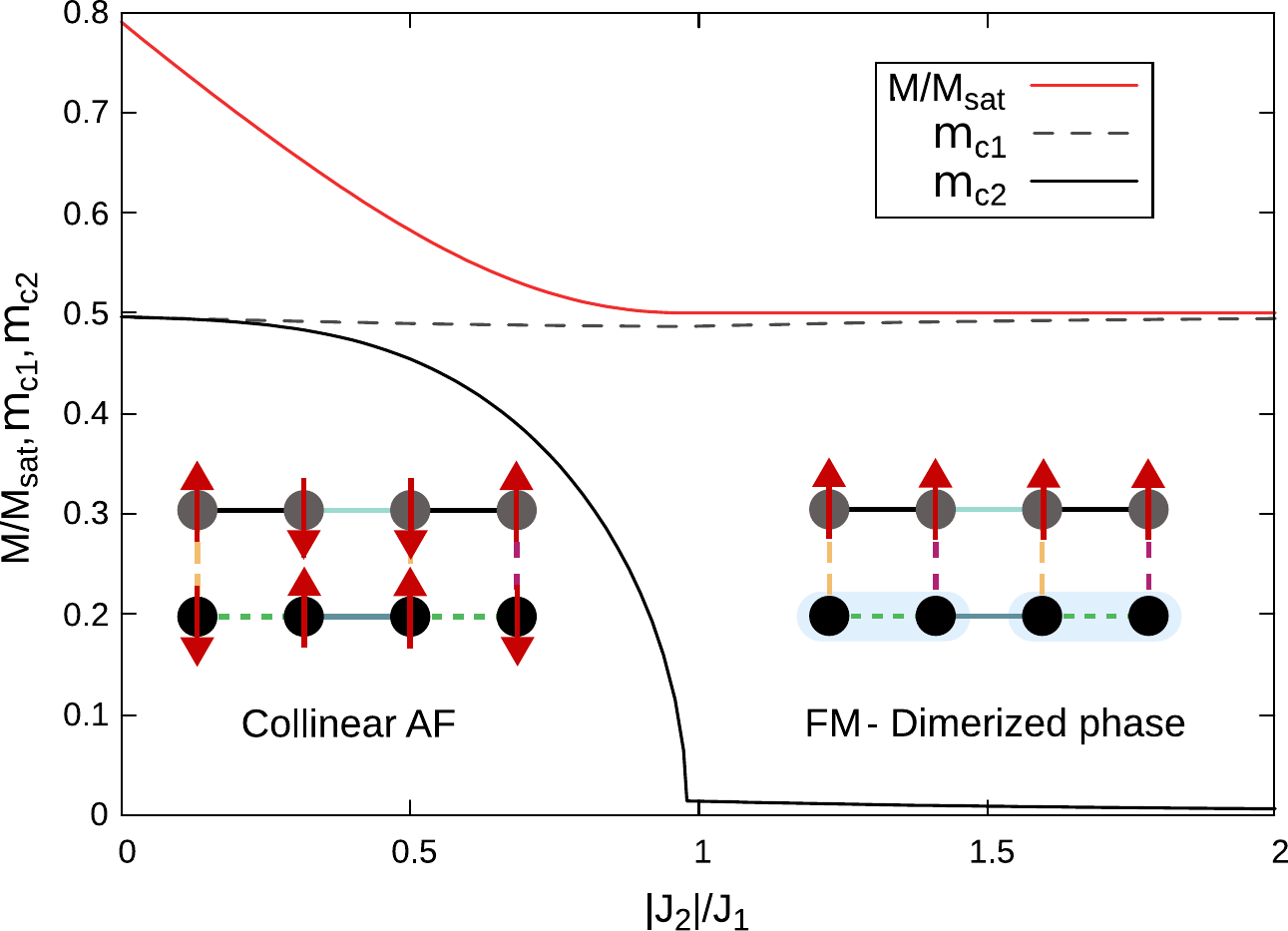}
    \caption{Total magnetization and average local moments of both quantum chains as a function of AF interaction $|J_{2}|/J_{1}$ at magnetic field $h^{z}/J_{1}=1.0$. The inset shows the  FM-dimerized phases that arise for $|J_{2}|/J_{1}>1.0$.}
    \label{fig:mags_vs_J2}
\end{figure}

Given the relevance of $J_2$ to the field-induced  FM-dimerized phase, we present in Fig. \ref{fig:mags_vs_J2} the dependence of the total magnetization and the average local moments of both chains on the AF coupling $|J_{2}|/J_{1}$ for a constant $h^{z}/J_1=1.0$, where this phase can be found. 
As $|J_{2}|/J_{1}$ increases from zero, the total magnetization (red curve) smoothly decreases from  $M/M_{sat}=0.8$ to approximately $M/M_{sat}=1/2$ for $|J_{2}|/J_{1}\gtrsim1.0$, reflecting the occurrence of a one-half magnetization plateau state.
The average local moments reveal that this plateau is characterized by a polarization of chain 1 (gray dashed curve), whose magnetization remains nearly constant across the entire field range.  
In contrast, chain 2 (black curve) undergoes a continuous suppression of its average local moment, nearly vanishing as $|J_{2}|/J_{1}\rightarrow 1.0$. 
Beyond this point, chain 2 enters a nonmagnetic state, consistent with the formation of a dimer phase, as illustrated schematically in the inset. 
Therefore, the strength of the AF coupling $J_{2}$ drives the system into a coexisting  FM-dimerized phase, where chain 1 retains the long-range FM order and chain 2 forms singlet pairs.  
Our analysis suggests that fine-tuning of $J_{n}$ allows control over the magnetic nature of the system under an external magnetic field, mainly adjusting the degree of frustration.
This mechanism may supply pathways for engineering exotic magnetic phases in other verdazyl-based compounds.
\begin{figure}[!t]
    \centering
    \includegraphics[width=0.48\textwidth]{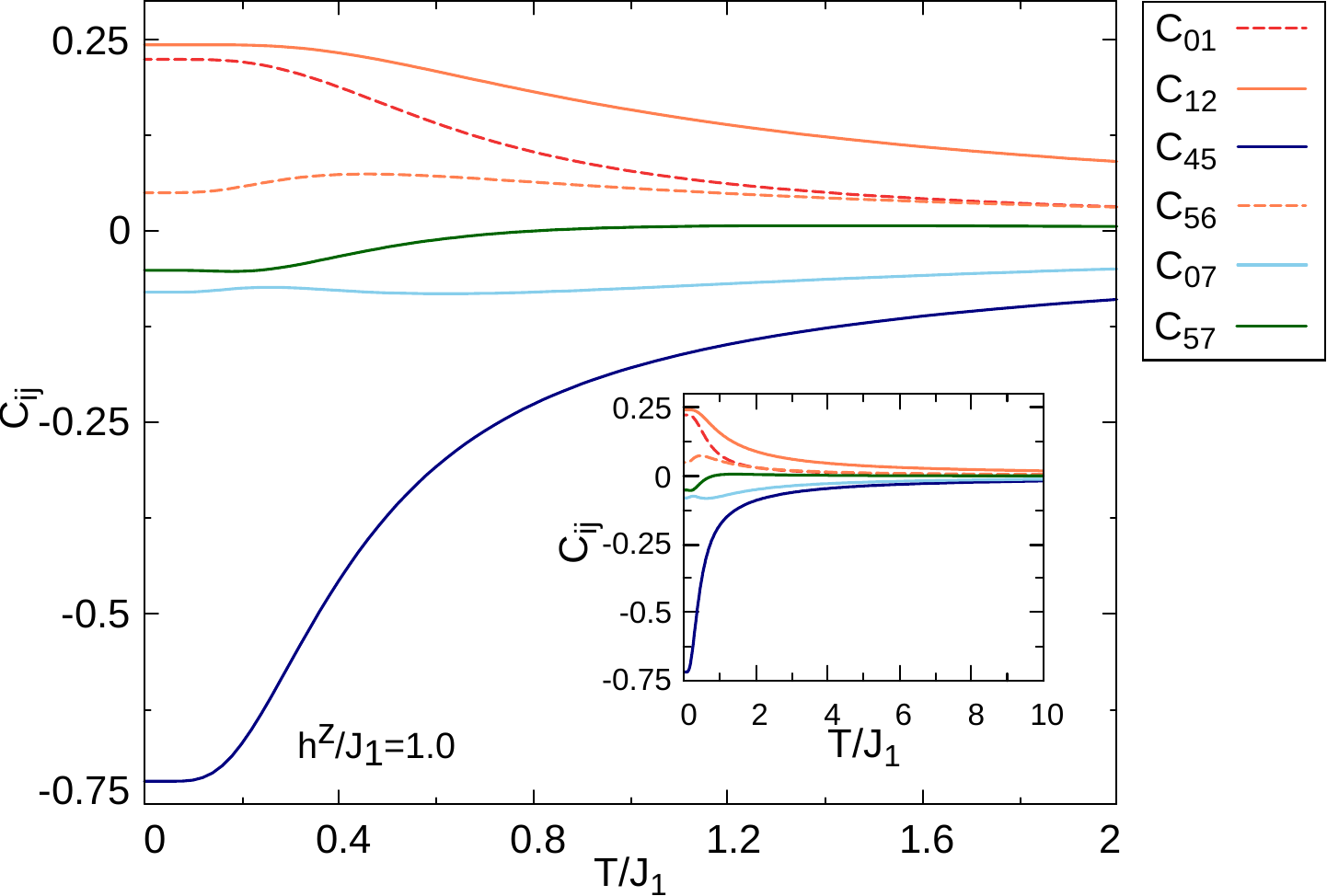}
    \caption{Spin-spin correlations $C_{ij}$ as a function of temperature within the eight-site CMF approach at constant field $h^{z}/J_{1}=1.0$. Some correlations have been omitted from the figure as they are nearly equivalent to those already shown, e.g., the correlations $C_{01}\approx C_{23}$, $C_{45}\approx C_{67}$, $C_{46}\approx C_{57}$ and the interchain correlations $C_{07}\approx C_{16} \approx C_{25} \approx C_{34}$.}
    \label{fig:correlations_vs_t}
\end{figure}
\subsection{Finite temperature correlations}
The temperature dependence of the spin-spin correlation functions under an applied magnetic field $h^{z}/J_1=1.0$ provides further insight into the stability of the field-induced phases. 
As shown in Fig. \ref{fig:correlations_vs_t}, the correlations of chain 1 ($C_{01}$, $C_{12}$,$C_{23}$) reflect its  FM polarization along the field direction, remaining close to 0.25 at low temperatures and decaying to zero as the temperature increases.
In contrast, the correlations $C_{45}$ and $C_{56}$ in chain 2 exhibit a different behavior at low temperatures, where $C_{45}$ approaches the perfect singlet value of \textminus0.75, and $C_{56}$ is significantly suppressed, consistent with the emergence of a dimer phase in chain 2. 
As the temperature increases, these correlations in chain 2 progressively weaken, signaling thermal suppression of the singlet states. In agreement with the previous discussion, the interchain correlations ($C_{07}$, $C_{16}$,...) and the correlation $C_{57}$ remain negligible throughout the temperature range. 
These results indicate that, under a range of applied fields, the system can stabilize a coexisting  FM-dimerized phase at low temperatures. 

\section{Summary and conclusion}\label{sec:summary}
We employed the CMF method to investigate the spin-1/2 Heisenberg model on an anisotropic square lattice inspired by the verdazyl-based compound ($o$-MePy-V)PF$_6$. This model features six distinct  FM and AF nearest-neighbor exchange interactions within two AF coupled quantum spin chains: one FM and the other  FM-AF.
We explored slightly modified combinations of these interactions to investigate the effect of varying degrees of frustration on the magnetic properties of the model.
Our discussion focuses on local magnetization patterns, average local moments, and spin–spin correlation functions under magnetic fields.

In the ground state, the local magnetization patterns recover the results obtained for the reference compound, 
exhibiting a collinear AF order in the transverse plane characterized by a twofold periodic structure \cite{Yamaguchi,lucas}.  
Small variations in the exchange couplings relative to the reference values can also drive a similar AF collinear order at zero field. However, these variations can give rise to a nontrivial magnetic behavior under an applied magnetic field.  
This becomes evident when the degree of frustration is increased by strengthening the AF character of the  FM-AF chain. 
In the presence of an external field, a nonmagnetic state emerges as a consequence of quantum fluctuations induced by the field in this chain, characterized by the vanishing of both longitudinal and transverse local magnetizations, while the other chain develops a FM order.  
As a result, a one-half magnetization plateau can be found in a range of applied fields. 
This means that small changes in the coupling parameters of the reference system can change the magnetization profile from a plateau-like regime to a well-defined magnetization plateau.
More importantly, the spin–spin correlation functions reveal the formation of spin-singlet pairs in the one-half magnetization plateau regime. The  FM-AF chain exhibits both the absence of local magnetic order and the presence of uncorrelated spin-singlet pairs, indicating a dimerized phase. Therefore, the ground state of the system with an enhanced degree of frustration can include a coexisting  FM-dimerized phase induced by an external field. 
This coexistence can also persist for low temperatures.

To conclude, the interplay between frustration and field-induced quantum fluctuations can drive the stabilization of exotic behaviors, such as the  FM-dimerized phase.
Remarkably, the degree of frustration required to obtain the FM-dimerized phase in such systems can be accessed by tuning the spin anisotropies of the reference compound~\cite{lucas}, or by adjusting the exchange parameters, as demonstrated in the present work.
These findings not only deepen our understanding of frustration-driven quantum phenomena but also offer valuable guidance for engineering novel magnetic phases in verdazyl-based compounds and other strongly correlated quantum magnets.

\section*{Acknowledgment}\label{sec:acknowledgment}
This work was supported by the Brazilian agencies Conselho Nacional de Desenvolvimento Científico e Tecnológico (CNPq), Grant Nos. 165330/2023-6 and 309652/2023-5, and Coordenação de Aperfeiçoamento de Pessoal de Nível Superior (Capes). FMZ and LMR also acknowledge support from the Fundação de Apoio ao Desenvolvimento do Ensino, Ciência e Tecnologia do Estado de Mato Grosso do Sul (Fundect).

\section*{Data availability }

The data that support the findings of this study are available from the corresponding author upon reasonable request.

\bibliographystyle{elsarticle-num-names}
\bibliography{bib.bib}% Produces the bibliography via BibTeX.

@article{PhysRevB.93.060407,
  title = {Spin-$S$ kagome quantum antiferromagnets in a field with tensor networks},
  author = {Picot, Thibaut and Ziegler, Marc and Or\'us, Rom\'an and Poilblanc, Didier},
  journal = {Phys. Rev. B},
  volume = {93},
  issue = {6},
  pages = {060407},
  numpages = {5},
  year = {2016},
  month = {Feb},
  publisher = {American Physical Society},
  doi = {10.1103/PhysRevB.93.060407},
  url = {https://link.aps.org/doi/10.1103/PhysRevB.93.060407}
}

@article{10.1063/1.2186278,
    author = {Moessner, Roderich and Ramirez, Arthur P.},
    title = {Geometrical frustration},
    journal = {Physics Today},
    volume = {59},
    number = {2},
    pages = {24-29},
    year = {2006},
    month = {02},
    issn = {0031-9228},
    doi = {10.1063/1.2186278},
    url = {https://doi.org/10.1063/1.2186278},
}

@article{Mydosh_2015,
doi = {10.1088/0034-4885/78/5/052501},
url = {https://dx.doi.org/10.1088/0034-4885/78/5/052501},
year = {2015},
month = {apr},
publisher = {IOP Publishing},
volume = {78},
number = {5},
pages = {052501},
author = {Mydosh, J A},
title = {Spin glasses: redux: an updated experimental/materials
                    survey},
journal = {Reports on Progress in Physics}
}

@article{LIU20221034,
title = {Gapless quantum spin liquid and global phase diagram of the spin-1/2 {$J_1-J_2$} square antiferromagnetic Heisenberg model},
journal = {Science Bulletin},
volume = {67},
number = {10},
pages = {1034-1041},
year = {2022},
issn = {2095-9273},
doi = {https://doi.org/10.1016/j.scib.2022.03.010},
url = {https://www.sciencedirect.com/science/article/pii/S2095927322001001},
author = {Wen-Yuan Liu and Shou-Shu Gong and Yu-Bin Li and Didier Poilblanc and Wei-Qiang Chen and Zheng-Cheng Gu}
}

@article{10.1093/ptep/ptae061,
    author = {Li, Shang-Wei and Jiang, Fu-Jiun},
    title = {A Comprehensive Study of the Phase Transitions of the Frustrated J1-J2 Ising Model on the Square Lattice},
    journal = {Progress of Theoretical and Experimental Physics},
    volume = {2024},
    number = {5},
    pages = {053A06},
    year = {2024},
    month = {04},
    issn = {2050-3911},
    doi = {10.1093/ptep/ptae061},
    url = {https://doi.org/10.1093/ptep/ptae061},
}

@article{lucas,
  title = {Interplay of frustration and quantum fluctuations in a spin-1/2 anisotropic square lattice},
  author = {Ramos, L. M. and Zimmer, F. M. and Schmidt, M.},
  journal = {Phys. Rev. B},
  volume = {112},
  issue = {1},
  pages = {014402},
  numpages = {9},
  year = {2025},
  month = {Jul},
  publisher = {American Physical Society},
  doi = {10.1103/svsc-bvjx},
  url = {https://link.aps.org/doi/10.1103/svsc-bvjx}
}

@article{Saito,
  title = {Phase diagram of the quantum spin-$\frac{1}{2}$ Heisenberg-$\mathrm{\ensuremath{\Gamma}}$ model on a frustrated zigzag chain},
  author = {Saito, Hidehiro and Hotta, Chisa},
  journal = {Phys. Rev. B},
  volume = {110},
  issue = {2},
  pages = {024409},
  numpages = {18},
  year = {2024},
  month = {Jul},
  publisher = {American Physical Society},
  doi = {10.1103/PhysRevB.110.024409},
  url = {https://link.aps.org/doi/10.1103/PhysRevB.110.024409}
}

@article{almeidabibiano,
  title = {Mixed-spin Heisenberg ladders in a magnetic field},
  author = {Almeida, D. S. and Bibiano-Filho, A. S. and da Silva, W. M. and Montenegro-Filho, R. R.},
  journal = {Phys. Rev. E},
  volume = {111},
  issue = {1},
  pages = {014149},
  numpages = {9},
  year = {2025},
  month = {Jan},
  publisher = {American Physical Society},
  doi = {10.1103/PhysRevE.111.014149},
  url = {https://link.aps.org/doi/10.1103/PhysRevE.111.014149}
}

@article{montenegro,
  title = {Ground-state phase diagram and thermodynamics of coupled trimer chains},
  author = {Montenegro-Filho, R. R. and Silva-J\'unior, E. J. P. and Coutinho-Filho, M. D.},
  journal = {Phys. Rev. B},
  volume = {105},
  issue = {13},
  pages = {134423},
  numpages = {10},
  year = {2022},
  month = {Apr},
  publisher = {American Physical Society},
  doi = {10.1103/PhysRevB.105.134423},
  url = {https://link.aps.org/doi/10.1103/PhysRevB.105.134423}
}

@article{wessel2017efficient,
  title={{Efficient Quantum Monte Carlo simulations of highly frustrated magnets: the frustrated spin-1/2 ladder}},
	author={Stefan Wessel and B. Normand and Frédéric Mila and Andreas Honecker},
	journal={SciPost Phys.},
	volume={3},
	pages={005},
	year={2017},
	publisher={SciPost},
	doi={10.21468/SciPostPhys.3.1.005},
	url={https://scipost.org/10.21468/SciPostPhys.3.1.005},
}

@article{Liu,
  title = {Existence of dimerized phases in frustrated spin ladder models},
  author = {Liu, Guang-Hua and Wang, Hai-Long and Tian, Guang-Shan},
  journal = {Phys. Rev. B},
  volume = {77},
  issue = {21},
  pages = {214418},
  numpages = {8},
  year = {2008},
  month = {Jun},
  publisher = {American Physical Society},
  doi = {10.1103/PhysRevB.77.214418},
  url = {https://link.aps.org/doi/10.1103/PhysRevB.77.214418}
}

@article{Metavitsiadis,
  title = {Spin-liquid versus dimer phases in an anisotropic ${J}_{1}$-${J}_{2}$ frustrated square antiferromagnet},
  author = {Metavitsiadis, Alexandros and Sellmann, Daniel and Eggert, Sebastian},
  journal = {Phys. Rev. B},
  volume = {89},
  issue = {24},
  pages = {241104},
  numpages = {5},
  year = {2014},
  month = {Jun},
  publisher = {American Physical Society},
  doi = {10.1103/PhysRevB.89.241104},
  url = {https://link.aps.org/doi/10.1103/PhysRevB.89.241104}
}

@article{Jiang,
  title = {Spin liquid ground state of the spin-$\frac{1}{2}$ square ${J}_{1}$-${J}_{2}$ Heisenberg model},
  author = {Jiang, Hong-Chen and Yao, Hong and Balents, Leon},
  journal = {Phys. Rev. B},
  volume = {86},
  issue = {2},
  pages = {024424},
  numpages = {14},
  year = {2012},
  month = {Jul},
  publisher = {American Physical Society},
  doi = {10.1103/PhysRevB.86.024424},
  url = {https://link.aps.org/doi/10.1103/PhysRevB.86.024424}
}

@article{pikulski2020two,
   title={Two coupled chains are simpler than one: field-induced chirality in a frustrated spin ladder},
   volume={10},
   ISSN={2045-2322},
   number={1},
   journal={Scientific Reports},
   author={Pikulski, Marek and Shiroka, Toni and Casola, Francesco and Reyes, Arneil P. and Kuhns, Philip L. and Wang, Shuang and Ott, Hans-Rudolf and Mesot, Joël},
   year={2020},
   doi={10.1038/s41598-020-72215-z},
   url={https://doi.org/10.1038/s41598-020-72215-z},
}

@article{Savary_2017,
doi = {10.1088/0034-4885/80/1/016502},
url = {https://dx.doi.org/10.1088/0034-4885/80/1/016502},
year = {2016},
month = {nov},
publisher = {IOP Publishing},
volume = {80},
number = {1},
pages = {016502},
author = {Savary, Lucile and Balents, Leon},
title = {Quantum spin liquids: a review},
journal = {Reports on Progress in Physics}
}

@article{Zhou,
  title = {Quantum spin liquid states},
  author = {Zhou, Yi and Kanoda, Kazushi and Ng, Tai-Kai},
  journal = {Rev. Mod. Phys.},
  volume = {89},
  issue = {2},
  pages = {025003},
  numpages = {50},
  year = {2017},
  month = {Apr},
  publisher = {American Physical Society},
  doi = {10.1103/RevModPhys.89.025003},
  url = {https://link.aps.org/doi/10.1103/RevModPhys.89.025003}
}

@article{karlov,
  title = {Breakdown of intermediate one-half magnetization plateau of spin-1/2 Ising-Heisenberg and Heisenberg branched chains at triple and Kosterlitz-Thouless critical points},
  author = {Karl'ov\'a, Katar\'{\i}na and Stre\ifmmode \check{c}\else \v{c}\fi{}ka, Jozef and Lyra, Marcelo L.},
  journal = {Phys. Rev. E},
  volume = {100},
  issue = {4},
  pages = {042127},
  numpages = {12},
  year = {2019},
  month = {Oct},
  publisher = {American Physical Society},
  doi = {10.1103/PhysRevE.100.042127},
  url = {https://link.aps.org/doi/10.1103/PhysRevE.100.042127}
}

@article{Song,
  title = {Entanglement entropy of the two-dimensional Heisenberg antiferromagnet},
  author = {Song, H. Francis and Laflorencie, Nicolas and Rachel, Stephan and Le Hur, Karyn},
  journal = {Phys. Rev. B},
  volume = {83},
  issue = {22},
  pages = {224410},
  numpages = {7},
  year = {2011},
  month = {Jun},
  publisher = {American Physical Society},
  doi = {10.1103/PhysRevB.83.224410},
  url = {https://link.aps.org/doi/10.1103/PhysRevB.83.224410}
}

@article{Laflorencie,
  title = {Spin-wave approach for entanglement entropies of the ${J}_{1}\ensuremath{-}{J}_{2}$ Heisenberg antiferromagnet on the square lattice},
  author = {Laflorencie, Nicolas and Luitz, David J. and Alet, Fabien},
  journal = {Phys. Rev. B},
  volume = {92},
  issue = {11},
  pages = {115126},
  numpages = {14},
  year = {2015},
  month = {Sep},
  publisher = {American Physical Society},
  doi = {10.1103/PhysRevB.92.115126},
  url = {https://link.aps.org/doi/10.1103/PhysRevB.92.115126}
}

@article{Udono,
  title = {Triplons, triplon pairs, and dynamical symmetries in laser-driven Shastry-Sutherland magnets},
  author = {Udono, Mina and Sato, Masahiro},
  journal = {Phys. Rev. B},
  volume = {110},
  issue = {20},
  pages = {L201112},
  numpages = {9},
  year = {2024},
  month = {Nov},
  publisher = {American Physical Society},
  doi = {10.1103/PhysRevB.110.L201112},
  url = {https://link.aps.org/doi/10.1103/PhysRevB.110.L201112}
}

@article{Honecker,
  title = {Multi-triplet bound states and finite-temperature dynamics in highly frustrated quantum spin ladders},
  author = {Honecker, Andreas and Mila, Fr\'ed\'eric and Normand, B.},
  journal = {Phys. Rev. B},
  volume = {94},
  issue = {9},
  pages = {094402},
  numpages = {21},
  year = {2016},
  month = {Sep},
  publisher = {American Physical Society},
  doi = {10.1103/PhysRevB.94.094402},
  url = {https://link.aps.org/doi/10.1103/PhysRevB.94.094402}
}

@article{Shao,
  title = {Nearly Deconfined Spinon Excitations in the Square-Lattice Spin-$1/2$ Heisenberg Antiferromagnet},
  author = {Shao, Hui and Qin, Yan Qi and Capponi, Sylvain and Chesi, Stefano and Meng, Zi Yang and Sandvik, Anders W.},
  journal = {Phys. Rev. X},
  volume = {7},
  issue = {4},
  pages = {041072},
  numpages = {26},
  year = {2017},
  month = {Dec},
  publisher = {American Physical Society},
  doi = {10.1103/PhysRevX.7.041072},
  url = {https://link.aps.org/doi/10.1103/PhysRevX.7.041072}
}

@article{Tran,
  title = {Spinon excitations in the quasi-one-dimensional $S=\frac{1}{2}$ chain compound $\mathrm{C}{\mathrm{s}}_{4}\mathrm{CuS}{\mathrm{b}}_{2}\mathrm{C}{\mathrm{l}}_{12}$},
  author = {Tran, Thao T. and Pocs, Chris A. and Zhang, Yubo and Winiarski, Michal J. and Sun, Jianwei and Lee, Minhyea and McQueen, Tyrel M.},
  journal = {Phys. Rev. B},
  volume = {101},
  issue = {23},
  pages = {235107},
  numpages = {8},
  year = {2020},
  month = {Jun},
  publisher = {American Physical Society},
  doi = {10.1103/PhysRevB.101.235107},
  url = {https://link.aps.org/doi/10.1103/PhysRevB.101.235107}
}

@article{LiHZ,
  title = {Dimerization-enhanced exotic magnetization plateau and magnetoelectric phase diagrams in skew-chain ${\mathrm{Co}}_{2}{\mathrm{V}}_{2}{\mathrm{O}}_{7}$},
  author = {Li, Z. H. and Han, X. T. and Dong, C. and Wang, H. W. and He, Z. Z. and Chen, R. and Liu, W. X. and Lu, C. L. and Kohama, Y. and Tokunaga, M. and Kindo, K. and Ouyang, Z. W. and Wang, J. F. and Yang, M.},
  journal = {Phys. Rev. B},
  volume = {109},
  issue = {9},
  pages = {094432},
  numpages = {8},
  year = {2024},
  month = {Mar},
  publisher = {American Physical Society},
  doi = {10.1103/PhysRevB.109.094432},
  url = {https://link.aps.org/doi/10.1103/PhysRevB.109.094432}
}

@article{yamamoto2017,
  title = {Exact diagonalization and cluster mean-field study of triangular-lattice XXZ antiferromagnets near saturation},
  author = {Yamamoto, Daisuke and Ueda, Hiroshi and Danshita, Ippei and Marmorini, Giacomo and Momoi, Tsutomu and Shimokawa, Tokuro},
  journal = {Phys. Rev. B},
  volume = {96},
  issue = {1},
  pages = {014431},
  numpages = {12},
  year = {2017},
  month = {Jul},
  publisher = {American Physical Society},
  doi = {10.1103/PhysRevB.96.014431},
  url = {https://link.aps.org/doi/10.1103/PhysRevB.96.014431}
}

@article{wang2025,
  title = {Experimental realization of antiferromagnetic Ising ground state on the triangular lattice},
  author = {Wang, Ke and Liu, Xing-Jian and Tu, Li-Ming and Zhang, Jia-Jie and Gladilin, Vladimir N. and Ge, Jun-Yi},
  journal = {Phys. Rev. B},
  volume = {111},
  issue = {22},
  pages = {224418},
  numpages = {8},
  year = {2025},
  month = {Jun},
  publisher = {American Physical Society},
  doi = {10.1103/syny-gvt3},
  url = {https://link.aps.org/doi/10.1103/syny-gvt3}
}

@article{Zimmer2014,
  title = {Interplay between spin-glass clusters and geometrical frustration},
  author = {Zimmer, F. M. and Silva, C. F. and Magalhaes, S. G. and Lacroix, C.},
  journal = {Phys. Rev. E},
  volume = {89},
  issue = {2},
  pages = {022120},
  numpages = {6},
  year = {2014},
  month = {Feb},
  publisher = {American Physical Society},
  doi = {10.1103/PhysRevE.89.022120},
  url = {https://link.aps.org/doi/10.1103/PhysRevE.89.022120}
}

@article{Schmidt_2017,
doi = {10.1088/1361-648X/aa6060},
url = {https://dx.doi.org/10.1088/1361-648X/aa6060},
year = {2017},
month = {mar},
publisher = {IOP Publishing},
volume = {29},
number = {16},
pages = {165801},
author = {Schmidt, M and Zimmer, F M and Magalhaes, S G},
title = {Spin liquid and infinitesimal-disorder-driven cluster spin glass in the kagome lattice},
journal = {Journal of Physics: Condensed Matter}
}

@article{Eric,
  title = {Cluster-Glass Phase in Pyrochlore $XY$ Antiferromagnets with Quenched Disorder},
  author = {Andrade, Eric C. and Hoyos, Jos\'e A. and Rachel, Stephan and Vojta, Matthias},
  journal = {Phys. Rev. Lett.},
  volume = {120},
  issue = {9},
  pages = {097204},
  numpages = {6},
  year = {2018},
  month = {Mar},
  publisher = {American Physical Society},
  doi = {10.1103/PhysRevLett.120.097204},
  url = {https://link.aps.org/doi/10.1103/PhysRevLett.120.097204}
}

@article{yamamoto2014,
  title = {Quantum Phase Diagram of the Triangular-Lattice $XXZ$ Model in a Magnetic Field},
  author = {Yamamoto, Daisuke and Marmorini, Giacomo and Danshita, Ippei},
  journal = {Phys. Rev. Lett.},
  volume = {112},
  issue = {12},
  pages = {127203},
  numpages = {5},
  year = {2014},
  month = {Mar},
  publisher = {American Physical Society},
  doi = {10.1103/PhysRevLett.112.127203},
  url = {https://link.aps.org/doi/10.1103/PhysRevLett.112.127203}
}

@article{yamamoto2015,
  title = {Microscopic Model Calculations for the Magnetization Process of Layered Triangular-Lattice Quantum Antiferromagnets},
  author = {Yamamoto, Daisuke and Marmorini, Giacomo and Danshita, Ippei},
  journal = {Phys. Rev. Lett.},
  volume = {114},
  issue = {2},
  pages = {027201},
  numpages = {5},
  year = {2015},
  month = {Jan},
  publisher = {American Physical Society},
  doi = {10.1103/PhysRevLett.114.027201},
  url = {https://link.aps.org/doi/10.1103/PhysRevLett.114.027201}
}

@article{Yamaguchi2021,
  title = {Gapped ground state in a spin-$\frac{1}{2}$ frustrated square lattice},
  author = {Yamaguchi, H. and Uemoto, N. and Okubo, T. and Kono, Y. and Kittaka, S. and Sakakibara, T. and Yajima, T. and Shimono, S. and Iwasaki, Y. and Hosokoshi, Y.},
  journal = {Phys. Rev. B},
  volume = {104},
  issue = {6},
  pages = {L060411},
  numpages = {5},
  year = {2021},
  month = {Aug},
  publisher = {American Physical Society},
  doi = {10.1103/PhysRevB.104.L060411},
  url = {https://link.aps.org/doi/10.1103/PhysRevB.104.L060411}
}

@article{Yamaguchi,
  title = {Field-enhanced quantum fluctuation in an $S=\frac{1}{2}$ frustrated square lattice},
  author = {Yamaguchi, H. and Sasaki, Y. and Okubo, T. and Yoshida, M. and Kida, T. and Hagiwara, M. and Kono, Y. and Kittaka, S. and Sakakibara, T. and Takigawa, M. and Iwasaki, Y. and Hosokoshi, Y.},
  journal = {Phys. Rev. B},
  volume = {98},
  issue = {9},
  pages = {094402},
  numpages = {6},
  year = {2018},
  month = {Sep},
  publisher = {American Physical Society},
  doi = {10.1103/PhysRevB.98.094402},
  url = {https://link.aps.org/doi/10.1103/PhysRevB.98.094402}
}

@article{kellermann2019quantum,
  title = {Quantum Ising model on the frustrated square lattice},
  author = {Kellermann, N. and Schmidt, M. and Zimmer, F. M.},
  journal = {Phys. Rev. E},
  volume = {99},
  issue = {1},
  pages = {012134},
  numpages = {6},
  year = {2019},
  month = {Jan},
  publisher = {American Physical Society},
  doi = {10.1103/PhysRevE.99.012134},
  url = {https://link.aps.org/doi/10.1103/PhysRevE.99.012134}
}

@article{Yang,
  title = {Quantum criticality and spin liquid phase in the Shastry-Sutherland model},
  author = {Yang, Jianwei and Sandvik, Anders W. and Wang, Ling},
  journal = {Phys. Rev. B},
  volume = {105},
  issue = {6},
  pages = {L060409},
  numpages = {7},
  year = {2022},
  month = {Feb},
  publisher = {American Physical Society},
  doi = {10.1103/PhysRevB.105.L060409},
  url = {https://link.aps.org/doi/10.1103/PhysRevB.105.L060409}
}

@article{Xi,
  title = {Plaquette valence bond solid to antiferromagnet transition and deconfined quantum critical point of the Shastry-Sutherland model},
  author = {Xi, Ning and Chen, Hongyu and Xie, Z. Y. and Yu, Rong},
  journal = {Phys. Rev. B},
  volume = {107},
  issue = {22},
  pages = {L220408},
  numpages = {6},
  year = {2023},
  month = {Jun},
  publisher = {American Physical Society},
  doi = {10.1103/PhysRevB.107.L220408},
  url = {https://link.aps.org/doi/10.1103/PhysRevB.107.L220408}
}

@article{PhysRevB.103.L220407,
  title = {Quantum critical phenomena in a spin-$\frac{1}{2}$ frustrated square lattice with spatial anisotropy},
  author = {Yamaguchi, H. and Iwasaki, Y. and Kono, Y. and Okubo, T. and Miyamoto, S. and Hosokoshi, Y. and Matsuo, A. and Sakakibara, T. and Kida, T. and Hagiwara, M.},
  journal = {Phys. Rev. B},
  volume = {103},
  issue = {22},
  pages = {L220407},
  numpages = {5},
  year = {2021},
  month = {Jun},
  publisher = {American Physical Society},
  doi = {10.1103/PhysRevB.103.L220407},
  url = {https://link.aps.org/doi/10.1103/PhysRevB.103.L220407}
}

@article{Singhania,
  title = {Cluster mean-field study of the {Heisenberg} model for {${\mathrm{CuInVO}}_{5}$}},
  author = {Singhania, Ayushi and Kumar, Sanjeev},
  journal = {Phys. Rev. B},
  volume = {98},
  issue = {10},
  pages = {104429},
  numpages = {9},
  year = {2018},
  month = {Sep},
  publisher = {American Physical Society},
  doi = {10.1103/PhysRevB.98.104429},
  url = {https://link.aps.org/doi/10.1103/PhysRevB.98.104429}
}

\end{document}